\documentclass[preprint,authoryear]{elsarticle}

\usepackage{commonm}
\usepackage{hyperref}
\usepackage{lineno}

\journal{ }

\begin{document}

\begin{frontmatter}

\title{Subgrid scale modeling of droplet bag breakup in VOF simulations}

\author[cornell]{Austin Han\corref{mycorrespondingauthor}}
\ead{ah2262@cornell.edu}
\author[cornell]{Olivier Desjardins}

\cortext[mycorrespondingauthor]{Corresponding author}

\address[cornell]{Sibley School of Mechanical and Aerospace Engineering, Cornell University, Ithaca, NY 14853, United States of America}

\begin{abstract}
    The mesh-dependency of the breakup of liquid films, including their breakup length scales and resulting drop size distributions, has long been an obstacle inhibiting the computational modeling of large-scale spray systems. With the aim of overcoming this barrier, this work presents a framework for the prediction and modeling of subgrid-thickness liquid film formation and breakup within two-phase simulations using the volume of fluid method. A two-plane interface reconstruction is used to capture the development of liquid films as their thickness decreases below the mesh size. The breakup of the film is predicted with a semi-analytical model that incorporates the film geometry captured through the two-plane reconstruction. The framework is validated against experiments of the bag breakup of a liquid drop at $\We=13.8$ through the comparison of the resulting drop size and velocity distributions. The generated distributions show good agreement with experimental results for drop resolutions as low as 25.6 cells across the initial diameter. The presented framework enables these drop breakup simulations to be performed at a computational cost three orders of magnitude lower than the cost of simulations utilizing adaptive mesh refinement.
\end{abstract}

\begin{keyword}
drop breakup\sep volume of fluid\sep liquid sheet breakup \sep atomization \sep subgrid scale modeling \sep drop size distribution
\end{keyword}

\end{frontmatter}


\section{Introduction}
This work proposes a multiscale modeling strategy in which the formation and evolution of liquid sheets is captured at subgrid thicknesses using the two-plane interface reconstruction method of \cite{hanCapturingThinStructures2024}, while the breakup of the sheets into droplets is predicted with a model adapted from the work of \cite{jackiwPredictionDropletSize2022}. The proposed sheet breakup model assumes that the rate of droplet formation is controlled by the Taylor--Culick speed \citep{taylorDynamicsThinSheets1959a,culickCommentsRupturedSoap1960} of a retracting sheet edge.
The preemptive transfer of liquid sheet volume from the Eulerian field to droplets in a Lagrangian point representation is similar to the method of \cite{kimSubgridscaleCapillaryBreakup2020} for the breakup of thin ligaments.
The use of a two-plane interface reconstruction within a Eulerian volume of fluid framework allows for the seamless transfer of volume from codimension-0 liquid structures to codimension-1 films, and the addition of a sheet breakup model facilitates the transfer of volume from codimension-1 films to codimension-3 Lagrangian point droplets. Liquid sheets are identified in the Eulerian grid with a connected components labeling (CCL) algorithm \citep{He2017}.
To demonstrate the proposed framework, it is applied to the aerodynamic fragmentation of a liquid drop in the bag breakup regime.

The rest of this paper is organized as follows: Section \ref{sec:configuration} describes the simulation configuration and physical parameters, while Section \ref{sec:numericalmethod} presents the numerical methods used in the present simulations. The film breakup model and its numerical implementation are described in Section \ref{sec:film}. Finally, Section \ref{sec:dropstats} presents a validation study of the droplet statistics.

\section{Computational methods}\label{sec:method}
\subsection{Physical configuration and parameters}\label{sec:configuration}
A stationary liquid drop of diameter $d_0$, density $\rho_l$, and viscosity $\mu_l$ is placed in a rectangular cuboid domain of size $(L_x,L_y,L_z)=(20 d_0,10 d_0,10 d_0)$ filled with an initially quiescent gas of density $\rho_g$ and viscosity $\mu_g$. The center of the drop is positioned $2d_0$ away from the lower $x$ domain boundary and is equidistant from the $y$ and $z$ boundaries. A Neumann outflow velocity boundary condition is prescribed to the upper $x$ domain boundary, while periodic boundary conditions are prescribed to the $y$ and $z$ domain boundaries. The lower $x$ domain boundary contains a Dirichlet inflow velocity condition.

The physical parameters of the problem are chosen to match the ethanol-air configuration of the experiments of \cite{guildenbecherCharacterizationDropAerodynamic2017} for their $\We=13.8$ case and are given in Table \ref{tab:physical_parameters}. Since $\Oh<10^{-2}$, the influence of viscosity on the drop deformation is weak relative to those of the competing inertia and surface tension forces  expressed through the Weber number \citep{hsiangNearlimitDropDeformation1992,tangBagFilmBreakup2023}. The gas inflow is generated from a precursor simulation of homogeneous isotropic turbulence with $\text{R}_{\lambda}=45$, and it has a turbulence intensity of $0.019\%$.
\begin{table*}[hbt]
    \centering
    \caption{Nondimensional parameters for the bag breakup simulations.}
    \begin{tabular}{c c c c c}
        $\Re=\rho_g U_0 d_0/\mu_g$ & $\We=\rho_g U_0^2 d_0/\sigma$ & $\rho_l/\rho_g$ & $\mu_l/\mu_g$ & $\Oh=\mu_l/\sqrt{\rho_l d_0 \sigma}$ \\ \hline
        1778 & 13.8 & 657.5 & 66.7 & 0.00543
    \end{tabular}
    \label{tab:physical_parameters}
\end{table*}
\subsection{Numerical method}\label{sec:numericalmethod}
The simulations in this work solve the Navier--Stokes equations for two-phase, incompressible, and immiscible flows of Newtonian fluids, given by
\begin{align}
    \nabla\cdot\bm{u}&=0, \\
    \frac{\partial\rho\bm{u}}{\partial t}+\nabla\cdot\left(\rho\bm{u}\otimes\bm{u}\right)&=-\nabla p+\nabla\cdot\left(\mu \left[\nabla\bm{u}+\nabla\bm{u}^{\intercal} \right]\right)+\sigma\kappa\delta_{\Gamma}\bm{n},
\end{align}
where $\rho$ is the fluid density, $\bm{u}$ is the velocity vector, $p$ is the pressure, and $\mu$ is  the fluid viscosity. The surface tension force is added on the right-hand side of the momentum equation with $\sigma$, $\kappa$, $\delta_{\Gamma}$, and $\bm{n}$ representing the surface tension coefficient, interface curvature, interface indicator function, and interface normal vector, respectively. 
The volume of fluid (VOF) method is used to implicitly represent the liquid-gas interface. The liquid volume fraction $\alpha$ is defined in each computational cell $\Omega_i$ as a local fraction of volume
\begin{equation}
    \alpha_i=\frac{1}{\mathcal{V}_{\Omega_i}}\int_{\Omega_i}\chi(\bm{x})dV,
\end{equation}
where $\mathcal{V}_{\Omega}$ is the cell volume, $\bm{x}\in\bbR^3$, and $\chi$ is an indicator function that follows
\begin{equation}
    \chi(\bm{x})=
    \begin{cases}
    1 & \text{for } \bm{x}\in \text{liquid} \\
    0 & \text{for } \bm{x}\in \text{gas}.
    \end{cases}
\end{equation}
The volume fraction $\alpha$ is transported with the equation
\begin{equation}\label{eq:voftransport}
    \frac{\partial\alpha}{\partial t}+\bm{u}\cdot\nabla\alpha=0,
\end{equation}
and the fluid density and viscosity in each cell are computed as an arithmetic mean of the liquid and gas values weighted by their respective volume fractions,
\begin{align}
    \rho&=\rho_l\alpha+\rho_g(1-\alpha), \label{eq:mixture_density}\\
    \mu &=\mu_l \alpha+\mu_g (1-\alpha). \label{eq:mixture_viscosity}
\end{align}

The Navier--Stokes and VOF advection equations are solved with the open-source, finite-volume flow solver NGA2 \citep{Desjardins2008,desjardinsDesjardiNGA2Objectoriented}, in which the advection of the volume fraction field is performed with the unsplit geometric method of \cite{Owkes2014}. The liquid-gas interface in each cell is approximated by one or two planes \citep{youngsTimedependentMultimaterialFlow1982}, and the plane orientations and positions are determined by the R2P method \citep{hanCapturingThinStructures2024}. The Interface Reconstruction Library (IRL) \citep{chiodiGeneralRobustEfficient2022,chiodiInterfaceReconstructionLibrary2023} is utilized for the geometric operations in advection and interface reconstruction. The surface tension force is computed using the continuum surface force model \citep{Brackbill1992,Popinet2009} with modifications to recover the pressure jump across subgrid-thickness films \citep{hanCapturingThinStructures2024}. The interface curvature for the surface tension force is computed by the volumetric fitting of paraboloids to the interface polygons \citep{Jibben2019,hanComparisonMethodsCurvature2024}. The eddy viscosity model of \cite{vremanEddyviscositySubgridscaleModel2004} provides a closure for the subgrid-scale turbulent stresses. The computational domain is discretized by a uniform Cartesian mesh with cubic cells. A computational cell size of $d_0/\Delta=25.6$ is used with a timestep size of $\Delta t U_0/D_0=2\times 10^{-2}$. In the experiments of \cite{guildenbecherCharacterizationDropAerodynamic2017}, the drop diameter for the $\We=13.8$ case is \qty{2.54}{\mm}, and the cell size is therefore \qty{99}{\um}, which is orders of magnitude larger than any film breakup length scale (on the order of a tenth to a few microns \citep{opferDropletairCollisionDynamics2014}).
All simulations are performed on a Beowulf cluster with nodes containing dual six-core Intel X5670 processors. The simulations have a computational cost of approximately 2800 core-hours. 

\section{Film breakup model}\label{sec:film}
After a film is perforated, the resulting film edge experiences a high surface tension force due to the large curvature. The surface tension force causes the film edge to recede away from the center of the perforation at a steady-state speed
\begin{equation}\label{eq:taylorculick}
    U_{TC}=\sqrt{\frac{2\sigma}{\rho_l h}},
\end{equation}
known as the Taylor--Culick speed \citep{taylorDynamicsThinSheets1959a,culickCommentsRupturedSoap1960}. As the film edge retracts, it collects liquid mass and forms a rim whose diameter is larger than the film thickness. Droplets are formed from two mechanisms: the growth of ligaments from the rim which continuously shed droplets \citep{agbaglahLongitudinalInstabilityLiquid2013,wangUnsteadySheetFragmentation2018}, and the collision of multiple rims \citep{neelFinesCollisionLiquid2020,agbaglahBreakupThinLiquid2021}.
The resulting droplet size distribution from a film breakup event can be approximated as a gamma distribution with number probability density function
\begin{equation}\label{eq:gammapdf}
    p_n(x=d/d_0;k,\theta)=\frac{x^{k-1}e^{-x/\theta}}{\Gamma(k)\theta^k},
\end{equation}
following the model of \cite{jackiwPredictionDropletSize2022}, where $k$ is the shape parameter, $\theta$ is the scale parameter, and $\Gamma$ is the gamma function. 

\subsection{Numerical implementation}
To identify constitutive cells of liquid films to which a film breakup model is applied, a connected components labeling (CCL) algorithm \citep{He2017} is used to identify films in the Eulerian volume fraction field. A CCL algorithm identifies contiguous regions within an array of binary values and outputs a list of cell indices for each contiguous region. The binary array used for film identification is
\begin{equation}
    X_i=
    \begin{cases}
        1 & \text{for } \bm{x_i}\in \mathcal{L} \\
        0 & \text{for } \bm{x_i}\notin \mathcal{L},
    \end{cases}
\end{equation}
where $\mathcal{L}$ is the set of liquid film cells as defined in \cite{hanCapturingThinStructures2024}.  
In each film cell $\Omega_i \in \mathcal{L}$, the cell-local film thickness is approximated as
\begin{equation}
    h_i=\frac{2\sum_{\Omega_j \in \mathscr{N}_i}\alpha_j V_j}{\sum_{\Omega_j \in \mathscr{N}_i}A_j},
    \label{eq:thickness}
\end{equation}
where $\mathscr{N}_i$ is the set of all cells that share a vertex with $\Omega_i$, $V_i$ is the cell volume, and $A_i$ is the surface area of the liquid-gas interface in the cell computed from the interface polygons.
For each identified film, the minimum film thickness $h_{\min}$ is the minimum of the cell-local thicknesses $h_i$ among the constituent film cells.
The film breakup is initiated when $h_{\min}<h_{threshold}$, where $h_{threshold}=10^{-3}d_0$ was chosen to match the minimum thickness of $h_{\min}=\qty{2.3}{\um}$ observed by \cite{jackiwPredictionDropletSize2022}. A drop undergoing breakup in the bag morphology will only produce one contiguous liquid film, but the proposed framework is capable of identifying and breaking up multiple films within the domain simultaneously in cases such as drop breakup in the multibag morphology and pressure swirl sprays \citep{chenDropletSizeDistribution2022}.

\emph{Hole initiation:} The procedure for producing Lagrangian droplets from the film breakup starts with the puncturing of the film at the point of minimum thickness when $h_{\min}<h_{threshold}$. At this location, the liquid volume from the ten cells with the lowest film thicknesses is removed and used to create Lagrangian droplets.
The indices of these ten cells are sorted and traversed by increasing thickness to replicate the physical process wherein the film is punctured at the point of minimum thickness, and the resulting hole expands from the thinnest film region to the thicker film regions near the rim.
As this list of cells is traversed, liquid volume from the cells is accumulated as a running sum $V_{sum}$, and the gamma distribution is recomputed at each cell. Simultaneously, a drop diameter $d_{drop}$ is sampled from the gamma distribution, and a Lagrangian droplet is formed when the accumulated volume is sufficient to produce a drop with $d=d_{drop}$. A new diameter is then sampled after the creation of the Lagrangian droplet. 
The initial velocity of the Lagrangian droplet is equal to the fluid mixture velocity $\bm{u}$ at the cell in which the droplet is formed, while the initial position is the cell liquid centroid with a random perturbation of maximum magnitude $\Delta/2$ tangent to the film interface.
The remaining liquid volume after droplet creation is used for subsequent droplet creation steps during the film breakup. Additionally, a list of edge cells $\mathcal{E}$ is initialized, where the initial edge cells are the set of liquid-containing cells neighboring any of the liquid-containing cells drained to create the initial film perforation.

\emph{Film retraction:} After the film is perforated, liquid volume is accumulated from the edge cells $\mathcal{E}$ into $V_{sum}$. The breakup procedure accumulates $V_{sum}$ over multiple timesteps and sheds droplets from the edge cells until the film is entirely fragmented. Tracking the simultaneous breakup of multiple films therefore requires the maintenance of multiple running sums along with the ability to track the position of these films over time. However, tracking simultaneous film breakup events is unnecessary for predicting the single bag breakup of a drop and is therefore unnecessary for this work. The volumetric accumulation rate per length for each edge cell $\Omega_i\in\mathcal{E}$ is
\begin{equation}
    q_{in,i}=h_iU_{TC,i},
\end{equation}
where $h_i$ is computed by Eq.\ \eqref{eq:thickness}, and $U_{TC_i}$ is computed by Eq.\ \eqref{eq:taylorculick}.
The assumption is made that all volume accumulated by the receding edge is immediately shed:
\begin{equation}
    q_{out,i}=q_{in,i}.
\end{equation}
In other words, the shedding model does not attempt to track the thickness of the receding rim. A further assumption is made that the edge length of each edge cell is $l\approx\Delta$, leading to a volumetric contribution for each cell of
\begin{equation}
    V_{out,i}=\min \left(h_{i}U_{TC_i}\Delta\Delta t,\alpha_i\Delta^3 \right),
\end{equation}
where the contribution is limited by the cell volume fraction to enforce mass conservation.
The accumulated volume sum for timestep $n$ is then computed as
\begin{equation}
    V_{sum}^{n}=V_{sum}^{n-1}+\sum_{\Omega_i\in\mathcal{E}}V_{out,i}.
\end{equation}
Additionally, the liquid volume fraction in each edge cell must be reduced to account for the volume contribution:
\begin{equation}
    \alpha_i=\alpha_i-V_{out,i}/\Delta^3.
\end{equation}

Once the volume sum has been computed, the edge cells are traversed in order of increasing thickness. At each traversed cell, the gamma distribution is recomputed, a drop diameter $d_{drop}$ is sampled from the distribution, a Lagrangian droplet is formed, and the droplet volume is subtracted from $V_{sum}$. The edge cells are traversed until either $V_{sum}$ is insufficient for a newly sampled drop diameter, or a droplet has been injected at each edge cell. Any remaining volume in $V_{sum}$ is carried over to the next timestep. The initial position of the Lagrangian droplets is the cell liquid centroid with a random perturbation of maximum magnitude $\Delta/2$ tangent to the film interface. The initial velocity is the sum of the fluid mixture velocity $\bm{u}$ and the retraction velocity $-U_{TC}\bm{n}_e$ at the cell in which the droplet is formed, where the edge normal $\bm{n}_e$ is computed as
\begin{equation}
    \bm{n}_e=\frac{\bm{c}^l-\bm{\Tilde{c}}^l}{\|\bm{c}^l-\bm{\Tilde{c}}^l\|},
\end{equation}
with $\bm{c}^l$ representing the cell liquid centroid, and $\bm{\Tilde{c}}^l$ representing the liquid centroid of the cell neighborhood.

After the droplets have been injected, if the film containing the edge has been completely broken into Lagrangian droplets, then any remaining $V_{sum}$ is distributed amongst the most recently injected droplets. Otherwise, the list of edge cells is updated through adding all cells neighboring cells that were drained through the drop shedding step.

\subsection{Edge cell advection}
During each timestep $n$, the list of edge cells $\mathcal{E}^n$ needs to be updated after the advection of the volume fraction field $\alpha$ (Eq.\ \eqref{eq:voftransport}) in order to maintain correspondence with the physical edge position. A scalar field value $E$ is used to indicate the current edge position, where
\begin{equation}
    E^n_i=
    \begin{cases}
        1 & \text{for } \Omega_i \in \mathcal{E}^n \\
        0 & \text{for } \Omega_i \notin \mathcal{E}^n.
    \end{cases}
\end{equation}
A semi-Lagrangian method is then used to transport the edge indicator to the updated positions. For each interfacial cell $\Omega_i$, the liquid centroid position is projected backward in time to the position $\bm{x}(t-\Delta t)$ using an implicit second-order Runge-Kutta step with interpolated velocities. If the projected position is within $\Omega_i$, then the projection is extended until the projected position is within a neighboring cell $\Omega_j$. If $E^n_j=1$ within $\Omega_j$, then $E^{n+1}_i$ is also given a value of 1. To prevent the edge indicator from being excessively propagated, a refinement step is performed using an edge detection algorithm. The edge detection is a centroid-shift method similar to that of \cite{ahmedEdgeCornerDetection2018a}, wherein a larger Euclidean distance $\phi=\|\bm{c}^l-\bm{\Tilde{c}}^l\|/\Delta$ between the cell-local and volume-filtered liquid centroids $\bm{c}^l$ and $\bm{\Tilde{c}}^l$, respectively, indicates a greater likelihood of a cell to be located at an edge. All cells with $E=1$ after the initial projection step with zero cells within the cell neighborhood with $\phi$ greater than a cutoff value $\phi_c=0.1$ have their edge indicator value reset to $E=0$. Finally, the list of edge cells is updated as
\begin{equation}
    \mathcal{E}=\{\Omega_i \mid E_i=1\}.
\end{equation}
It is possible that newer methods based on deep neural networks \citep{yuECNetEdgeawarePoint2018,tanCoarsetofinePipeline3D2022} would increase the robustness of the edge detection, but this extension is left to future work.

\section{Droplet statistics}\label{sec:dropstats}
\subsection{Experimental validation}
This section presents a validation study of the proposed modeling framework against the experiments of \cite{guildenbecherCharacterizationDropAerodynamic2017} through comparison of the resulting drop size distributions. 
The bag breakup of a liquid drop produces droplets of three characteristic sizes \citep{chouTemporalPropertiesSecondary1998,guildenbecherCharacterizationDropAerodynamic2017}. The smallest of these sizes are of droplets produced from the fragmentation of the liquid sheet. The remaining toroidal rim has a nonuniform minor radius, as roughly cylindrical sections of the rim are connected to large ``nodes'' that grow from the rim during the early drop deformation process. The cylindrical sections of the rim break due to capillary instabilities, and the resulting droplets define the second of the characteristic sizes. After the breakup of the cylindrical sections of the rim, the nodes remain as detached, roughly spherical liquid structures and define the largest of the three characteristic sizes.
The drop sizes from the sheet fragmentation are predicted with the model discussed in Section \ref{sec:film}, while those from the breakup of the rim are predicted with a model similar to the model of \cite{kimSubgridscaleCapillaryBreakup2020}.
Since the characteristic sizes of the rim and node droplets are much larger than that of the film droplets, the number of rim and node droplets produced from a single breakup event is much smaller than the number of film droplets, and a number drop size distribution would not adequately capture the rim and node droplets, considering that their collective volume can be as much as 90\% of the initial drop volume \citep{guildenbecherCharacterizationDropAerodynamic2017}. Therefore, a volume-weighted drop size distribution is computed, where each drop is weighted by its volume.

To replicate the turbulent gas jet inflow in the experiments, a turbulent inflow velocity is used for the simulations. The random perturbations to the drop deformation from the turbulent velocity field have a significant impact on the eventual droplet distribution, and 45 realizations of the drop breakup simulation are therefore performed for ensemble averaging. The experimental drop size distribution from \cite{guildenbecherCharacterizationDropAerodynamic2017} is produced from 44 breakup realizations. 
The droplet size distribution also includes detached liquid structures in the Eulerian volume fraction field that have not been transferred to Lagrangian droplet representation through the film or ligament breakup models. The drop diameter of these structures is taken as the diameter of a sphere with equal volume to the structure, or $d_{detached}=(6V/\pi)^{1/3}$. These detached liquid structures include the previously discussed node droplets that remain from the ligament-mode rim breakup.

\begin{figure*}[htb]
    \centering\includegraphics[width=\textwidth]{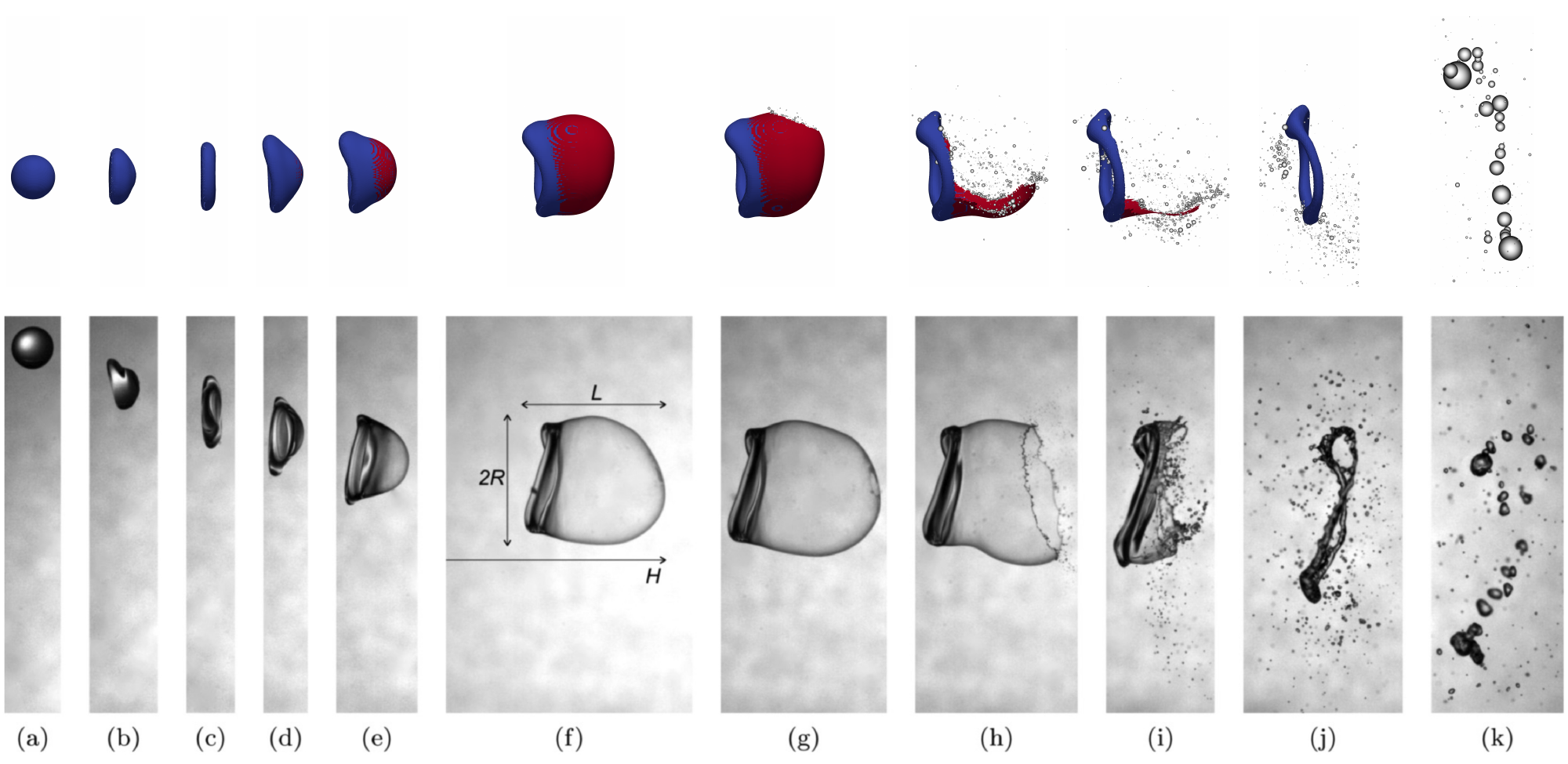}
    \caption{Time sequence of the bag breakup of a drop at $\We=13.8$ and $d_0/\Delta=25.6$ on the top row. Experimental images from \cite{opferDropletairCollisionDynamics2014} shown on the bottom, reproduced with permission from the publisher. Simulation images correspond to times $t/\tau=0$ (a), 0.58 (b), 0.97 (c), 1.91 (d), 2.14 (e), 2.42 (f), 2,46 (g), 2.57 (h), 2.61 (i), 2.69 (j), 3.63 (k). Experimental images correspond to times $t/\tau= 0$ (a), 0.53 (b), 0.92 (c), 1.11 (d), 1.30 (e), 1.47 (f), 1.50 (g), 1.53 (h), 1.67 (i), 1.83 (j), 2.17 (k). Single-plane reconstructions are shown in blue, while two-plane reconstructions are shown in red. Lagrangian droplets are represented by gray spheres. In the top row of (k), the remaining detached structures after the breakup of the film and rim are transferred to Lagrangian representation for the collection of dropsize statistics.}
    \label{fig:breakup-full}
\end{figure*}

Figure \ref{fig:breakup-full} shows a time sequence of the bag breakup for one of the 45 turbulent realizations alongside images from the experiments of \cite{opferDropletairCollisionDynamics2014}. In the simulation, as in the experiments, a hole is formed in the film, and droplets are shed from the boundary of the hole continuously as the hole expands until the hole collides with the toroidal rim in Fig. \ref{fig:breakup-full}(j).

\begin{figure}
    \centering
    \begin{tikzpicture}
        \begin{axis}[
            xlabel=$d/d_0$,
            ylabel=$f(d/d_0)$,
            xmin=0,xmax=0.1,ymin=0,
            x tick label style={
                /pgf/number format/.cd,
                fixed,
                fixed zerofill,
                precision=2,
                /tikz/.cd,
            },
            xticklabel shift={.1cm},
            legend style={at={(0.98,0.98)},anchor=north east,draw=none},
            legend cell align={left},
            ]
            \addplot+ [area legend,ybar interval,gray,fill=gray!50!white] table[x=edges,y=experiment] {\npdffilmsim};
            \addlegendentry{Experiment}
            \label{plot:npdfexp}
            \addplot [sixcolor1] table[x=D/D0,y=Time-resolved] {\npdffilmsim};
            \addlegendentry{Simulation}
            \label{plot:npdfretract}

        \end{axis}
    \end{tikzpicture}        

    \caption{Comparison of drop size number distribution of film droplets with the high-magnification measurements in Fig. 9 of \cite{guildenbecherCharacterizationDropAerodynamic2017}.}
    \label{fig:film_npdf}
\end{figure}
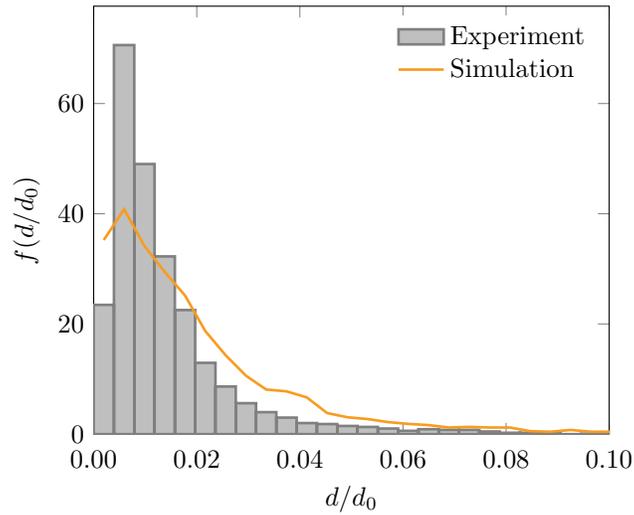

Figure \ref{fig:vpdfshedding} shows the volume-weighted drop size distribution from the sum of the 45 simulation breakup realizations alongside the \cite{guildenbecherCharacterizationDropAerodynamic2017} distribution. The distribution is multimodal with peaks at $d/d_0\approx 0.05,0.27$, and $0.5$, corresponding to the film breakup mode, the ligament breakup mode, and detached structures, respectively. The drop size distribution was found to not depend on mesh resolution for $d_0/\Delta>25$.
The location, width, and height of the distribution corresponding to the film droplets matches well that of the corresponding experimental distribution at $d/d_0\approx 0.05$.
Good experimental agreement is also observed in the number distribution in Fig. \ref{fig:film_npdf}, where only the drop sizes resulting from the film breakup for one of the 45 turbulent realizations are represented.
The results demonstrate that the two-plane R2P interface reconstruction is able to accurately predict the film volume and thickness distribution from the drop deformation and bag formation. 

Likewise, the central and right peaks, respectively corresponding to the ligament breakup mode and detached structures, have similar locations, widths, and heights to the corresponding experimental distributions.
A possible reason for the slight overprediction in ligament droplet sizes is that the collision of the receding film and the toroidal rim is not captured in the separate film and ligament breakup models. This collision imparts a disturbance force that influences the onset and wavelength of the capillary instabilities that ultimately fragment the rim. \cite{jackiwPredictionDropletSize2022} found that accounting for the collision perturbation decreased the average droplet size in their predictions of the rim fragmentation relative to a prediction that assumes that Rayleigh--Plateau instabilities are the sole influence.

\begin{figure}
    \centering
    \begin{subfigure}[b]{\columnwidth}
        \centering
        \begin{tikzpicture}
            \begin{axis}[
                xlabel=$d/d_0$,
                ylabel=$f_v(d/d_0)$,
                xmin=0,
                xmax=0.8,
                ymin=0,
                area legend,
                legend style={at={(0.02,0.98)},anchor=north west,draw=none,fill=none},
                ]
                \addplot+ [ybar interval,] [gray,fill=gray!50!white] table[x=edges,y=vals] {\vpdfexp};
                \label{plot:svpdfexp}
                \addplot+ [fill,no markers,opacity=0.5,solid] [sixcolor0] table[x=D/D0,y=film] {\vpdfshed} \closedcycle;
                \label{plot:sfilm}
                \addplot+ [fill,no markers,opacity=0.5,solid] [sixcolor1] table[x=D/D0,y=rim] {\vpdfshed} \closedcycle;
                \label{plot:srim}
                \addplot+ [fill,no markers,opacity=0.5,solid] [sixcolor2] table[x=D/D0,y=node] {\vpdfshed} \closedcycle;
                \label{plot:snode}
                \addplot [line legend,solid,line width=1.5pt] [sixcolor3] table[x=D/D0,y=sum] {\vpdfshed};
                \label{plot:ssum}
            \end{axis}
        \end{tikzpicture}        
    \end{subfigure}
    \caption{Volume-weighted drop size distribution measured from 45 breakup realizations using the film breakup model. The distribution of drop sizes produced from the film breakup is shaded in (\ref{plot:sfilm}), while that from the ligament breakup mechanism is shaded in (\ref{plot:srim}), and the size distribution of any remaining liquid structures is shown in (\ref{plot:snode}). The collective sum is plotted in (\ref{plot:ssum}). The measurements from \cite{guildenbecherCharacterizationDropAerodynamic2017} are shown in (\ref{plot:svpdfexp}).
    }
    \label{fig:vpdfshedding}
\end{figure}
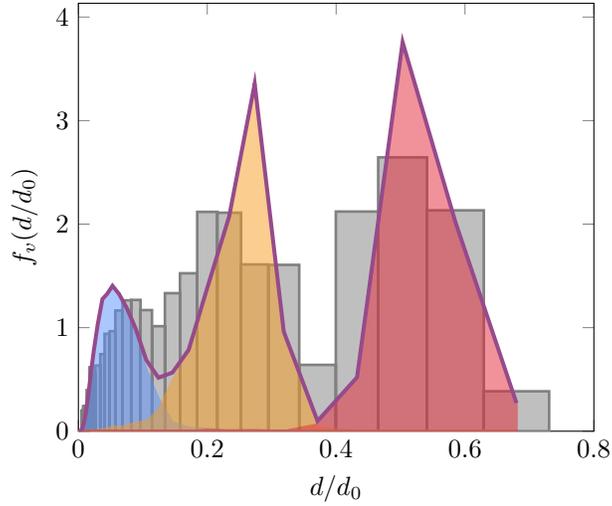

\begin{figure}[hbt]
    \centering
    \includegraphics[width=0.8\columnwidth]{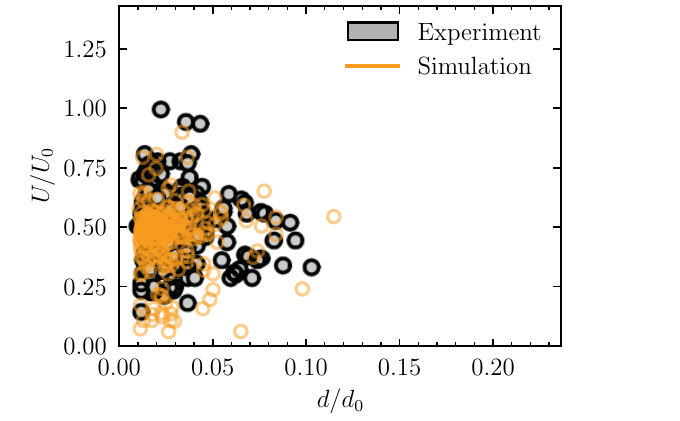}
    \caption{Scatter plot of film droplet diameter vs. velocity magnitude compared to the measurements of \cite{guildenbecherCharacterizationDropAerodynamic2017}, Fig. 7b. Each data set represents 250 randomly selected droplets.}
    \label{fig:velvsdiam}
\end{figure}

Finally, Fig. \ref{fig:velvsdiam} shows a scatter plot of sizes vs. velocity magnitudes of film droplets produced from the breakup model for one realization compared to data from the \cite{guildenbecherCharacterizationDropAerodynamic2017} experiments. Due to the addition of the Taylor--Culick retraction velocity to the initial droplet velocities, the distribution of velocities from the proposed model closely matches that of the experiment.

\section{Conclusions}\label{sec:conclusions}
A framework has been presented that combines a two-plane interface reconstruction and a semi-analytical breakup model within a volume of fluid simulation to predict the breakup of and resulting drop sizes from subgrid films. Due to the lack of mesh-induced breakup with the two-plane reconstruction, the breakup of the film, including the prediction of resulting drop sizes, is modeled, following \cite{jackiwPredictionDropletSize2022}, by assuming that the drop sizes follow a gamma distribution. The injection of Lagrangian droplets from the breakup of a film is spread between multiple timesteps by modeling the breakup as the shedding of droplets from a receding film rim with velocity governed by a Taylor--Culick law \citep{taylorDynamicsThinSheets1959a,culickCommentsRupturedSoap1960}. Thin ligaments undergo subgrid breakup using a capillary instability model similar in implementation to the work of \cite{kimSubgridscaleCapillaryBreakup2020}. The canonical test case of a liquid drop undergoing bag breakup at $\We=13.8$, using physical parameters from the experiments of \cite{guildenbecherCharacterizationDropAerodynamic2017}, is used to evaluate the proposed method. The film breakup model produces droplet size and velocity distributions that are in good agreement with the experimental results. The use of low mesh resolutions enables the prediction of bag breakup at computational costs on the order of $10^3$ core-hours, in contrast to simulations utilizing adaptive mesh refinement, which can incur costs on the order of $10^6$ core-hours \citep{lingDetailedNumericalInvestigation2023}.

Although the proposed methods have been demonstrated on the bag breakup of a single liquid drop, they are readily and affordably extensible to spray atomization simulations utilizing similar mesh resolutions of $d_0/\Delta=\order{10}$ with respect to the nozzle diameter \citep{vuComputationalStudyTwofluid2023}. The work of \cite{jackiwAerodynamicDropletAtomization2023} shows promise in the application of theoretical drop breakup models to the prediction of drop sizes from twin-fluid sprays with minimal modification to the former model. As the proposed framework makes no assumptions about the relative volumes or connectivity of ligaments and films in the domain, it should likewise be generalizable to the modeling of various spray applications with little modification.

\section{Acknowledgements}
This work was sponsored by the Office of Naval Research (ONR) as part of the Multidisciplinary University Research Initiatives (MURI) Program, under grant number N00014-16-1-2617.  The views and conclusions contained herein are those of the authors only and should not be interpreted as representing those of ONR, the U.S. Navy, or the U.S. Government. 

\bibliography{allreferences}

\end{document}